\documentclass[12pt,a4paper]{article}
\usepackage[applemac]{inputenc}
\usepackage{amssymb}
\usepackage{amsmath}
\usepackage{amsopn}
\newcommand{\LL}{\mathbb{L}}
\newcommand{\CC}{\mathbb{C}}

\newcommand{\MM}{\mathbb{M}}
\newcommand{\NN}{\mathbb{N}}

\newcommand{\RR}{\mathbb{R}}

\newcommand{\frAr}{\mathfrak{A}(\rr)}

\newcommand{\frb}{\mathfrak{B}}

\newcommand{\kaa}{\mathcal{A}}
\newcommand{\kbb}{\mathcal{B}}

\newcommand{\kcc}{\mathcal{C}}

\newcommand{\kf}{\mathcal{F}}
\newcommand{\kh}{\mathcal{H}}

\newcommand{\kL}{\mathcal{L}}
\newcommand{\mm}{\mathcal{M}}

\newcommand{\kP}{\mathcal{P}}
\newcommand{\kQ}{\mathcal{Q}}
\newcommand{\kR}{\mathcal{R}}
\newcommand{\kS}{\mathcal{S}}
\newcommand{\kU}{\mathcal{U}}

\newcommand{\kT}{\mathcal{T}}

\newcommand{\gb}{\beta}

\newcommand{\eps}{\varepsilon}
\newcommand{\gga}{\gamma}

\newcommand{\gl}{\lambda}

\newcommand{\gs}{\sigma}

\newcommand{\gt}{\tau}
\newcommand{\gtt}{\theta}
\newcommand{\gz}{\zeta}
\newcommand{\tm}{\subseteq}

\newtheorem{definition}{Definition}[section]
\newtheorem{proposition}{Proposition}[section]
\newtheorem{theorem}{Theorem}[section]
\newtheorem{lemma}{Lemma}[section]
\newtheorem{corollary}{Corollary}[section]
\newtheorem{remark}{Remark}[section]

\newcommand{\pr}{\kP(\kR)}

\newcommand{\pa}{\kP(\kaa)}

\newcommand{\pc}{\kP(\kcc)}

\newcommand{\qr}{\mathcal{Q}(\mathcal{R})}
\newcommand{\qm}{\kQ(\kT(M))}

\newcommand{\qpr}{\mathcal{Q}_{P}(\mathcal{R})}

\newcommand{\qa}{\mathcal{Q}(\mathcal{A})}

\newcommand{\qc}{\mathcal{Q}(\mathcal{C})}
\newcommand{\qmm}{\mathcal{Q}(\mm)}

\newcommand{\ql}{\mathcal{Q}(\LL)}

\newcommand{\qal}{\mathcal{Q}_{a}(\LL)}
\newcommand{\qbl}{\kQ_{b}(\LL)}

\newcommand{\all}{\forall}
\newcommand{\ex}{\exists}
\newcommand{\rr}{\kR}

\newcommand{\el}{E_{\gl}}

\newcommand{\we}{\wedge}
\newcommand{\We}{\bigwedge}
\newcommand{\Ve}{\bigvee}

\newcommand{\tto}{\mapsto}
\newcommand{\lra}{\Longrightarrow}
\newcommand{\llra}{\Longleftrightarrow}
\newcommand{\smm}{\setminus}

\newcommand{\pp}{\perp}

\newcommand{\lir}{\gl \in \RR}

\newcommand{\nin}{n \in \NN}

\newcommand{\urb}{\overset{-1}}

\newcommand{\ccn}{\kcc^{n}}

\newcommand{\pbr}{\kP^{\gb}(\rr)}
\newcommand{\qbr}{\kQ^{\gb}(\rr)}

\newcommand{\mnc}{\MM_{n}(\kcc)}

\newcommand{\scc}{s_{\kcc}}
\newcommand{\ur}{\kU(\rr)}

\begin{document}

\begin{center}
    {\LARGE \bf Stone spectra of finite von Neumann algebras of type
    $\rm{I}_{n}$}\\
    ~\\
    {\large \bf Hans F. de Groote}\\
    ~\\
    May 2, 2006
\end{center}
    
\begin{abstract}
    \noindent{In} this paper, we clarify the structure of the Stone spectrum of
    an arbitrary finite von Neumann algebra $\rr$ of type
    $\rm{I}_{n}$. The main tool for this investigation is a
    generalized notion of rank for projections in von Neumann algebras
    of this type.
\end{abstract}

\section{Introduction}
\label{in}

The Stone spectrum $\ql$ of a lattice $\LL$ is the set of all
\emph{maximal dual ideals} in $\LL$, endowed with the following
topology: For $a \in \LL$, let 
\[
    \qal := \{ \frb \in \LL \mid a \in \frb \}.
\]
Then 
\[
    \kQ_{0}(\LL) = \emptyset, \ \kQ_{1}(\LL) = \LL \ \textit{and} \
    \qal \cap \qbl = \kQ_{a \we b}(\LL) \ \textit{for all} \ a, b \in 
    \LL.
\]
Here $0$ denotes the minimal and $1$ the maximal element of $\LL$. We 
assume here that such elements exist in $\LL$.\footnote{This is not an 
essential restriction. If $\LL$ is complete, they exist anyway: $0 =
\We_{a \in \LL}a$ and $1 = \Ve_{a \in \LL}a$.} Hence the sets $\qal , 
\ a \in \LL$, form a basis of a topology for $\ql$. We call $\ql$,
together with this topology, the Stone spectrum of the lattice $\LL$.
This is an obvious generalization of Stone's construction
(\cite{stone}). Stone's motivation was to represent an abstract
Boolean algebra by a Boolean algebra of sets. However, there are two
other scenarios, quite different from Stone's, that lead to the same
construction.\\
~\\
If $\kS$ is a presheaf on a complete lattice $\LL$, we
can define \emph{germs} of $\kS$ in every \emph{filter base} in $\LL$,
i.e. in every subset $\kf$ that satisfies
\begin{enumerate}
    \item  [(i)] $0 \notin \kf$,

    \item  [(ii)] if $a, b \in \kf$, there is $c \in \kf$ such that $c
    \leq a, b$,
\end{enumerate}
as elements of the inductive limit 
\[
\lim_{\overset{\longrightarrow}{a \in \kf}}\kS(a).
\]
Of course, if $a \in \LL \smm \{0\}$, $\{a\}$ is a filter base, but in
general it makes no sense to regard $\kS(a)$ as a germ. Moreover, the 
set of \emph{all} filter bases is a vast object. For the definition of
an etale space corresponding to the presheaf $\kS$, it is therefore
meaningful to consider germs in maximal filter bases. 
This is why we speak of \emph{quasipoints} instead of maximal filter
bases. Now it is easy to see that maximal filter bases are 
nothing else but maximal dual ideals in $\LL$. The
\emph{sheafification} of $\kS$ is therefore defined over the Stone
spectrum of the lattice $\LL$ (\cite{deg3}).\\
~\\
The Stone spectrum $\qr$ of a von Neumann algebra $\rr$ is by definition 
the Stone spectrum of its projection lattice $\pr$. If $\rr$ is
abelian, $\qr$ is homeomorphic to the Gelfand spectrum of $\rr$
(\cite{deg3}). The Stone spectrum of $\rr$ is therefore a
generalization of the Gelfand spectrum to the non-abelian case.
Moreover, also the Gelfand transformation has a natural generalization
to the non-abelian case. If $A$ is a selfadjoint element of a von Neumann
algebra $\rr$, and if $E = (\el)_{\lir}$ is the spectral family of
$A$, we define
\[
    f_{A}(\frb) := \inf \{ \gl \mid \el \in \frb \}
\]
for all $\frb \in \qr$. The function $f_{A} : \qr \to \RR$, called the
\emph{observable function corresponding to $A$}, is
continuous and coincides with the Gelfand transform of $A$ if $\rr$ is
abelian (\cite{deg4}). This means that the Stone spectrum of $\rr$ is 
a natural generalization of the Gelfand spectrum.\\
The Gelfand spectrum of an abelian von Neumann algebra is, in general, a rather
wild object. So it is to be expected that the Stone spectrum of a
non-abelian von Neumann algebra can have a very intricate structure.\\
~\\
In the case of $\rr = \kL(\CC^{n})$ however, the situation is quite
simple: let $\frb$ be a quasipoint of $\rr$ and let $P_{0} \in \frb$
be a projection whose rank is minimal in the set $\{ rk(P) \mid P \in 
\frb \}$. Pick a subprojection $Q$ of $P_{0}$ that has rank one. Then 
$rk(P \we P_{0}) \leq rk( P_{0})$ and $P \we P_{0} \in \frb$ for all $P 
\in \frb$, hence $P_{0} \leq P$ for all $P \in \frb$. This implies $Q 
\leq P$ for all $P \in \frb$ and, therefore, $Q \in \frb$ by the
maximality of $\frb$. Hence 
\[
\frb = \{ P \in \pr \mid Q \leq P \}
\]
for a unique $Q \in \pr$ of rank one. Since in a factor the abelian
projections are minimal, the foregoing result can be expressed as:
each quasipoint contains an abelian projection. \\
~\\
If $\rr$ is a von Neumann algebra of type $\rm{I}_{n}, \ \nin$, with
center $\kcc$, the above argument is not applicable since the rank
of a projection is infinite in general. Nevertheless the result that
each quasipoint of $\rr$ contains an abelian projection is still
true\footnote{This result was already stated in \cite{doe1a}, but the 
proof contains an error: theorem 33 is not true in general (see remark
\ref{rk5})}.
Even the simple idea of the foregoing proof is transferable - provided
the notion of rank is suitably generalized. This generalization is the
basic idea of this paper. \\
~\\
Moreover, we investigate the topological structure of $\qr$. It is
shown that $\qr$ is a sheaf over the Stone spectrum $\qc$ of the center
$\kcc$ of $\rr$. The projection mapping of this sheaf is given by
\[
\begin{array}{cccc}
    \gz_{\kcc} : & \qr & \to & \qc\\
     & \frb & \tto & \frb \cap \kcc.
\end{array}
\]
This implies that $\qr$ is a locally compact space. The fibres of
$\gz_{\kcc}$ are discrete and representable as quotients of the
unitary group $\ur$ of $\rr$ modulo a subgroup (depending on the
fibre).\\
~\\
In section \ref{tr} we solve the ``trace-problem'' for quasipoints of 
finite von Neumann algebras $\rr$ of type $\rm{I}_{n}$. This problem
is the issue whether for a given quasipoint $\frb$ of a von Neumann
algebra  $\rr$ there is a maximal abelian von Neumann subalgebra $\mm$
of $\rr$ such that $\frb \cap \mm$ is a quasipoint of $\mm$. We
show that this is the case for all finite von Neumann algebras of type
$\rm{I}_{n}$. Moreover, we show that the property
\[
    \ex \ \frb \in \qr \ \all \ \kaa \in \frAr : \ \frb \cap \kaa \in
    \qa,
\]
where $\frAr$ denotes the set of all abelian von Neumann subalgebras
of $\rr$, is only satisfied in the trivial case that $\rr$ is abelian.

\section{Local action of partial isometries on Stone spectra}
\label{ac}

Let $\rr$ be a von Neumann algebra with center $\kcc$. 
The following result (\cite{deg3}, \cite{doe1a}) means that the Stone 
spectrum of the center $\kcc$ of an arbitrary von Neumann algebra
$\rr$ is a \emph{quotient} of $\qr$. For the sake of completeness, we 
present here also the proof given in \cite{deg3} for the first half of
this proposition.

\begin{proposition}\label{ac1}
    Let $\rr$ be a von Neumann algebra with center $\kcc$ and let
    $\kaa$ be a von Neumann subalgebra of $\kcc$. Then the mapping
    \[
	\gz_{\kaa} : \frb \tto \frb \cap \kaa
    \]
    is an open continuous, and therefore identifying, mapping from $\qr$
    onto $\qa$. Moreover
    \[
	\gz_{\kaa}(\frb) = \{ s_{\kaa}(P) \ | \ P \in \frb \}
    \]           
    for all $\frb \in \qr$, where 
    \[
	s_{\kaa}(P) := \We \{ Q \in \pa \ | \ P ≤ Q \}
    \]
     is the $\kaa$-support of $P \in \pr$. \\
     Conversely, if $\mm$ is a von Neumann subalgebra of $\rr$ such
     that $\frb \cap \mm \in \qmm$ for all $\frb \in \qr$, then $\mm$ 
     is contained in the center of $\rr$.
\end{proposition}
\emph{Proof:} $\frb \cap \kaa$ is clearly a dual ideal in $\pa$. Let 
$\gb \in \qa$ be a quasipoint that contains $\frb \cap \kaa$ and let
$C \in \gb$. If $C \notin \frb \cap \kaa$ then $C \notin \frb$. Hence 
there is some $P \in \frb$ such that $P \we C = 0$. Because $C$ is
central this means $PC = 0$. But then $P = PC + P(I - C) = P(I - C)$, 
i.e. $P ≤ I - C$. This implies $I - C \in \frb \cap \kaa \tm \gb$, a
contradiction to $C \in \gb$. Hence $\frb \cap \kaa$ is a quasipoint
in $\kaa$.\\
It follows immediately from the definition of the $\kaa$-support that
\[
    \all \ P, Q \in \pr : \ P ≤ s_{\kaa}(P) \ \ \text{and} \ \
    s_{\kaa}(P \we Q) ≤ s_{\kaa}(P) \we s_{\kaa}(Q)
\]
holds. This implies that $\{ s_{\kaa}(P) \ | \ P \in \frb \}$ is a
filter base contained in $\frb \cap \kaa$. Because of $s_{\kaa}(P) =
P$ for all $P \in \pa$, we must have equality. \\
Now we prove that
\begin{enumerate}
    \item  [(i)] $\all \ P \in \pr : \ \gz_{\kaa}(\qpr) =
    \kQ_{s_{\kaa}(P)}(\kaa)$ and

    \item  [(ii)] $\all \ Q \in \pa : \ \overset{-1}{\gz_{\kaa}}(\kQ_{Q}
    (\kaa)) = \kQ_{Q}(\rr)$
\end{enumerate}
hold: It is obvious that $\gz_{\kaa}(\qpr)$ is contained in
$\kQ_{s_{\kaa}(P)}(\kaa)$. Let $\gga \in \kQ_{s_{\kaa}(P)}(\kaa)$.
Then $P \in \urb{s_{\kaa}}(\gga)$, and we shall show that this implies
that $\{ PQ \mid Q \in \gga \} \cup \gga$ is a filter base in $\pr$.
Since $\gga$ consists of central projections, $\{ PQ \mid Q \in \gga \} 
\cup \gga$ is a filter base if and only if
\[
    \all \ Q \in \gga : \ PQ \ne 0.
\]
Assume that $PQ = 0$ for some $Q \in \gga$. Then $P ≤ I - Q$, hence
also $s_{\kaa}(P) ≤ I - Q$, contradicting $s_{\kaa}(P) \in \gga$.
Let $\frb$ be a quasipoint in $\pr$ that contains $\{PQ \mid Q \in
\gga \} \cup \gga$.
Because of $s_{\kaa}(Q) = Q$ for all $Q \in \gga$ we obtain
\[
    \gga = s_{\kaa}(\{PQ \mid Q \in \gga \} \cup \gga) \tm s_{\kaa}(\frb) =
    \gz_{\kaa}(\frb).
\]
Hence $\gga = \gz_{\kaa}(\frb)$ since $\gz_{\kaa}(\frb)$ and $\gga$
are quasipoints in $\pa$. This proves $(i)$. $(ii)$ follows from the
fact that each quasipoint in $\pa$ is contained in a quasipoint in
$\pr$. 
Properties $(i)$ and $(ii)$ imply that $\gz_{\kaa}$ is open, continuous
and surjective.\\
Now let $\mm$ be a von Neumann subalgebra of $\rr$ such that  $\frb
\cap \mm \in \qmm$ for all $\frb \in \qr$. Without loss of generality 
we can assume that $\mm = lin_{\CC}\{P, I \}$ for some nonzero
projection  $P \in \rr$. Then our condition is equivalent to 
\[
    \all \ \frb \in \qr : \ P \in \frb \ \textit{or} \ I - P \in \frb.
\]
Let $Q$ be an arbitrary nonzero projection in $\rr$. If $Q \we P + Q
\we (I - P) < Q$, there is a quasipoint $\frb$ of $\rr$ that contains 
$Q - (Q \we P + Q \we (I - P))$. Then $Q \in \frb$, but $Q \we P, \ Q 
\we (I - P) \notin \frb$. According to our condition we have $P \in
\frb$ or $I - P \in \frb$, hence $Q \we P \in \frb $ or $Q \we (I - P)
\in \frb$, a contradiction. Thus $Q = Q \we P + Q \we (I - P)$.
Therefore $Q \we P = (Q \we P)P = QP$, hence $PQ = QP$. Since $Q$ was 
arbitrary, $P$ is in the center of $\rr$.\ \ $\Box$ \\
~\\
If $\frb \in \qr$ and $F \in \frb$, the set
\[
    \frb_{F} := \{ P \in \frb \mid P ≤ F \}    
\]
is called the $F$-socle of $\frb$. It is easy to see that a quasipoint
 is uniquely determined by any of its socles.\\
~\\
Let $\gtt \in \rr$ be a partial isometry, i.e. $E := \gtt^\ast \gtt$ and
$F := \gtt \gtt^\ast$ are projections. $\gtt$ has kernel $E(\kh)^\pp$ and
maps $E(\kh)$ \emph{isometrically} onto $F(\kh)$. Now it is easy to
see that for any projection $P_{U} ≤ E$ we have
\begin{equation}
    \gtt P_{U} \gtt^\ast = P_{\gtt(U)}.
    \label{eq:vn1}
\end{equation}
A consequence of this relation is
\begin{equation}
    \all \ P, Q ≤ E : \ \gtt (P \we Q) \gtt^\ast = (\gtt P \gtt^\ast) \we 
	(\gtt Q \gtt^\ast). 
    \label{eq:vn2}
\end{equation}
If $\frb \in \kQ_{E}(\rr)$ then
\begin{equation}
    \gtt_{\ast}(\frb_{E}) := \{ \gtt P \gtt^\ast \ | \ P \in \frb_{E} \}
    \label{eq:vn3}
\end{equation}               
is the $F$-socle of a (uniquely determined) quasipoint $\gtt_{\ast}(\frb)
\in \kQ_{F}(\rr)$:\\
Equation \ref{eq:vn2} guarantees that
$\gtt_{\ast}(\frb_{E})$ is a filter base. Let $\Tilde{\frb}$ be a
quasipoint that contains $\gtt_{\ast}(\frb_{E})$. Then $\gtt_{\ast}(\frb_{E})
\tm \Tilde{\frb}_{F}$. Assume that this inclusion is proper. If $Q \in
\Tilde{\frb}_{F} \smm \gtt_{\ast}(\frb_{E})$ then $\gtt^\ast Q \gtt
\notin \frb_{E}$ and therefore there is some $P \in \frb_{E}$ such that
$P \we \gtt^\ast Q \gtt = 0$. But then $\gtt P \gtt^\ast \we Q = 0$, a
contradiction. This shows that we obtain a mapping
\[
    \begin{array}{cccc}
	\gtt_{\ast} : & \kQ_{\gtt^\ast \gtt}(\rr) & \to & \kQ_{\gtt
	\gtt^\ast}(\rr)  \\
	 & \frb & \tto & \gtt_{\ast}\frb,
     \end{array}
\]
where $\gtt_{\ast}\frb$ denotes the quasipoint determined by
$\gtt_{\ast}(\frb)$. \\
It is easy to see that $\gtt_{\ast}$ is a homeomorphism with inverse
$(\gtt^\ast) _{\ast}$. Note that $\gtt_{\ast}$ is globally defined if 
$\gtt$ is given by a unitary operator. 

\begin{definition}\label{ac2}
    Let $\gb \in \qc$. A quasipoint $\frb$ of $\rr$ is called a quasipoint
    over $\gb$ if $\frb \cap \kcc = \gb$ holds. Similarly, $P \in \pr$
   is called a projection over $\gb$ if $\scc(P) \in \gb$ holds. We denote
   by $\qbr$ the set of all quasipoints over $\gb$ and by $\pbr$ the
   set of all projections over $\gb$. 
\end{definition}

\begin{lemma}\label{ac3}
    Let $\gtt \in \rr$ be a partial isometry and $\frb \in
    \kQ_{\gtt^{\ast}\gtt}(\rr)$. Then $\frb \in \qbr$ if and only if
    $\gtt_{\ast}\frb \in \qbr$.
\end{lemma}
\emph{Proof:} Let $E := \gtt^{\ast}\gtt \in \frb$ and $F :=
\gtt\gtt^{\ast}$. Then $F = \gtt E \gtt^{\ast}$, $\scc(E) = \scc(F)$,
and the $\scc(E)$-socle $\gb_{\scc(E)} = \{ p\scc(E) \mid p \in \gb \}$
is mapped by conjugation with $\gtt$ onto $\gb_{\scc(E)} = \gtt
\gb_{\scc(E)} \gtt^{\ast} \tm \gga_{\scc(E)}$, where $\gga := \kcc
\cap \gtt_{\ast}\frb$. If $\gb_{\scc(E)} \ne \gga_{\scc(E)}$, there is
$q \in \gga$ such that $pq = 0$. But this contradicts the inclusion 
$\gb_{\scc(E)} \tm \gga_{\scc(E)}$. Hence $\gb_{\scc(E)} =
\gga_{\scc(E)}$, so $\gb = \gga$. \ \ $\Box$\\
~\\
The image $\gtt_{\ast}\frb$ of $\frb$ depends on the partial isometry 
$\gtt$, not only on the projection $\gtt^{\ast}\gtt \in \frb$. If, for
example, $\rr$ is a factor of type $\rm{III}$, every non-zero projection
$P \in \rr$ is equivalent to $I$, so $\qr \cong \qpr$. The situation
is considerably simpler for \emph{abelian} quasipoints.

\begin{definition}\label{ac4}
    A quasipoint of $\rr$ is called abelian if it contains an abelian 
    projection.
\end{definition}
The term ``abelian quasipoint'' is motivated by the following fact:
If $E \in \frb$ is an abelian projection then the $E$-socle $\frb_{E}$,
which determines $\frb$ uniquely, consists entirely
of abelian projections. Moreover every subprojection of an abelian
projection $E$ is of the form $CE$ with a suitable central projection $C$.
Hence
\[
    \frb_{E} = \{ CE \ | \ C \in \frb \cap \kcc \}
\]
if $E$ is abelian.\\
~\\
Let $\frb$ be an abelian quasipoint over $\gb \in \qc$ and let $E, F
\in \frb$ be abelian projections. Then
\[
    E \we F = pE = qF
\]
for suitable $p, q \in \gb$, so $pqE = pqF$. If $G$ is an abelian
projection, equivalent to $E$ via the partial isometry $\gtt$, then
$G \in \pbr$ and $pqE \sim pqG$ via the partial isometry $pq\gtt$.
Let $\frb'$ be a quasipoint over $\gb$ that contains $G$. If $H \in
\frb'$ is any abelian projection, $rG = rH$ for some $r \in \gb$.
Therefore, $pqrE \sim pqrH$. Since a quasipoint is determined by
any of its socles, it follows that $\frb' = \gtt_{\ast}\frb$ for every
partial isometry $\gtt \in \rr$ such that $\gtt^{\ast}\gtt \in \frb, \
\gtt\gtt^{\ast} \in \frb'$, and both are abelian. Summing up, we
have proved  the following result which already appears (with a
similar proof) in \cite{doe1a}:

\begin{proposition}\label{ac5}
    Let $\rr$ be a von Neumann algebra with center $\kcc$ and let
    $\frb, \frb' \in \qr$ be abelian quasipoints. Then $\frb' = \gtt_{\ast}
    \frb$ for some partial isometry $\gtt \in \rr$ if and only if
    $\frb \cap \kcc = \frb' \cap \kcc$. 
\end{proposition}

\section{Rank over a quasipoint of the center}
\label{rk}

Let $\rr$ be a von Neumann algebra of type $\rm{I}_{n} \ (\nin)$ and
let $\kcc$ be the center of $\rr$. If $\kcc = \CC$, then $\rr \cong
\kL(\CC^{n})$ and the rank $rk(P)$ of a projection $P \in \rr$ is the 
maximal number of pairwise orthogonal (equivalent) abelian
subprojections of $P$. In this case, it coincides with the rank of $P$
as a Hilbert space operator. In general, \emph{the rank of $P \in
\pr$ as a Hilbert space operator is infinite.} If we want to define the
rank of $P$ as a natural number, we can do this only
\emph{locally} over the quasipoints of $\kcc$. The definition emerges 
quite naturally from the following

\begin{theorem}\label{rk1}(\cite{kr2}, Corollary 6.5.5)\\
    If $P$ and $E$ are projections in a von Neumann algebra $\rr$ 
    with center $\kcc$, $s_{\kcc}(P) = s_{\kcc}(E)$, and $E$
    is abelian in $\rr$, there is a family $(p_{j})$ of central
    projections in $\rr$ with sum $s_{\kcc}(P)$ such that $p_{j}P$ is 
    the sum of $j$ equivalent abelian projections. If $\rr
    s_{\kcc}(E)$ is of type $\rm{I}_{n}$, then $1 ≤ j ≤ n$.
\end{theorem}
It is useful to throw a short look at the proof of this result to see 
how the projections$p_{j}$ depend on $P$.\\
~\\
Let $E \in \pr$ be an arbitrary abelian projection with central 
support $I$ and let $\gb \in \qc$. If $P \in \rr$ is a projection
over $\gb$, the algebra $\rr s_{\kcc}(P)$ is of type $\rm{I}_{n}$,
too, and $\scc(P)E$ is an abelian projection with central support
$\scc(P)$. Hence there are mutually orthogonal central projections
$p_{1}, \ldots, p_{n}$ with sum $\scc(P)$ such that $p_{j}P$ is the
sum of $j$ equivalent abelian projections $(1 ≤ j ≤ n)$. Since
$\scc(P) \in \gb$, there is a unique $j_{P} \in \{1,\ldots, n\}$ such 
that $p_{j_{P}} \in \gb$.

\begin{definition}\label{rk2}
    Let $P \in \rr$ be a projection over $\gb \in \qc$. Then
    \[
        rk_{\gb}(P) := j_{P}
    \]
    is called the rank of $P$ over $\gb$.
\end{definition}
Note that if $\rr$ is a factor, then $\qc = \{\{I\}\}$ and $rk_{\{I\}}
(P) = rk(P)$ for all nonzero $P \in \pr$.

\begin{proposition}\label{rk3}   
    The rank over $\gb$ has the following properties:
    \begin{enumerate}
        \item  [(i)] If $P \in \pbr$, then
	$rk_{\gb}(pP) = rk_{\gb}(P)$ for all $p \in \gb$.
	
        \item  [(ii)] $P \in \pbr$ has rank $n$ over $\gb$ if and only
	if there is a $p \in \gb$ such that $p ≤ P$.
        
        \item  [(iii)] If $P, Q \in \pbr$ and $P ≤ Q$, then
	$rk_{\gb}(P) ≤ rk_{\gb}(Q)$.
	
        \item  [(iv)] $P \in \pbr$ has rank $1$ over $\gb$ if and only
	if there is a $p \in \gb$ such that $pP$ is abelian.
    
        \item  [(v)] If $P \in \pr \smm\{0\}$, the function
	$\gb \tto rk_{\gb}(P)$ is defined and locally constant on
	$\kQ_{\scc(P)}(\kcc)$.
    \end{enumerate}
\end{proposition}
\emph{Proof:} $(i)$ follows directly from elementary results on the
equivalence of projections.\\
~\\
$(ii)$ \ If $rk_{\gb}(P) = n$, there are mutually orthogonal 
equivalent abelian projections $E_{1},\ldots, E_{n}$ and $p_{n} \in \gb$
such that $p_{n}P = E_{1} + \cdots + E_{n}$. Since $\rr$ is of type
$\rm{I}_{n}$, there are mutually orthogonal equivalent abelian projections
$F_{1}, \ldots, F_{n}$ such that $p_{n}\scc(P) = F_{1} + \cdots + F_{n}$. 
But $E_{1} + \cdots + E_{n} \sim F_{1} + \cdots + F_{n}$ by \cite{kr2},
proposition 6.2.2, and $p_{n}P ≤ p_{n}\scc(P)I_{n}$. Hence $p_{n}P =
p_{n}\scc(P) = p_{n}$, because $\rr$ is finite and $p_{n} ≤
\scc(P)$.\\
~\\
$(iii)$ \ Since $P ≤ Q$, $\scc(P) ≤ \scc(Q)$ and, therefore, $P ≤
\scc(P)Q$. Hence we may assume that $\scc(P) = \scc(Q)$. Now, by
definition, there are $p_{j}, q_{k} \in \gb$ and orthogonal families
$(E_{1},\ldots, E_{j}), \ (F_{1},\ldots, F_{k})$ such that $E_{1} \sim
\cdots \sim E_{j}, \ F_{1} \sim \cdots \sim F_{k}$ and
\begin{eqnarray*}
    p_{j}P & = & E_{1} + \cdots + E_{j},  \\
    q_{k}Q & = & F_{1} + \cdots + F_{k}.
\end{eqnarray*}
Since $\scc(E_{l}) = p_{j}$ and $\scc(F_{m}) = q_{k}$ for all $l ≤ j, 
\ m ≤ k$, $p_{j}q_{k}E_{l} \sim p_{j}q_{k}F_{m}$ for all $l ≤ j, \ m ≤
k$. If $k < j$, we deduce from
\begin{eqnarray*}
    p_{j}q_{k}P & = & p_{j}q_{k}E_{1} + \cdots + p_{j}q_{k}E_{j},  \\
    p_{j}q_{k}Q & = & p_{j}q_{k}F_{1} + \cdots + p_{j}q_{k}F_{k}
\end{eqnarray*}
that $p_{j}q_{k}Q$ is equivalent to a proper subprojection of
$p_{j}q_{k}P$, contradicting $p_{j}q_{k}P ≤ p_{j}q_{k}Q$.\\
~\\  
$(iv)$ follows from the very definition of the rank over $\gb$.\\
~\\ 
$(v)$ \ Clearly, $\gb \tto rk_{\gb}(P)$ is constant on
$\kQ_{p_{j}}(\kcc)$. \ \ $\Box$\\
~\\
Property $(i)$ shows that $rk_{\gb}(P)$ is merely an invariant of the
{\bf set}
\[
    [P]_{\gb} := \{ pP \mid p \in \gb \}.
\]
This leads to the following notion.

\begin{definition}\label{rk4}
     Let $\rr$ be an arbitrary von Neumann algebra with nontrivial center 
    $\kcc$, $\gb \in \qc$ and define
    \[
         \all \ P, Q \in \pr : \ (P \sim_{\gb} Q \ :\llra \ \ex \ p \in \gb : \
         pP = pQ).
    \]
\end{definition}
Clearly, since $\gb$ is a dual ideal, $\sim_{\gb}$ is an equivalence 
relation on $\pr$. We denote the equivalence class of $P \in \pr$ by
$[P]_{\gb}$, or simply by $[P]$, as long as the quasipoint $\gb$ is
fixed.

\begin{remark}\label{rk5}
     We can extend $\sim_{\gb}$ to all of $\rr$ by
    defining
    \[
	\all \ A, B \in \rr : \ (A \sim_{\gb} B \ :\llra \ \ex \ p \in
	\gb : \ pA = pB).
    \]
    The set $[\rr]$ of equivalence classes becomes a $\ast$- algebra by setting
    \[
	[A]^{\ast} := [A^{\ast}], \ [A] + [B] := [A + B] \ \ \textit{and} \ \
	[A][B] := [AB].
    \]
    For $A \in \rr$ let
    \[
	|[A]| := \inf \{ |pA| \mid p \in \gb \}.
    \]
    $|[A]|$ is called the $\gb$-seminorm of $[A]$. Indeed, the
    $\gb$-seminorm is a submultiplicative seminorm on $[\rr]$ that
    satisfies $|[A]^{\ast}[A]| = |[A]|^{2}$, but, if
    $\kcc$ is not trivial, it is not a norm. Moreover, it is easy to
    prove, using the spectral theorem, that
    \[
	|[a]| = |\gt_{\gb}(a)|,
    \]
    where $\gt_{\gb}$ is the character of $\kcc$ that is induced by
    $\gb$, holds for all $a \in \kcc$:\\
    It suffices to prove the assertion for $a ≥ 0$, since
    then $|b|_{\gb}^{2} = |b^{\ast}b|_{\gb} = |\gt_{\gb}(b^{\ast}b)| =
    |\gt_{\gb}(b)|^{2}$ for all $b \in \kcc$. \\
    Let $a ≥ 0$. Then for all $p \in \gb$ we have $|\gt_{\gb}(a)| =
    |\gt_{\gb}(pa)| ≤ |pa|$, hence
    \[
    |\gt_{\gb}(a)| ≤ |a|_{\gb}.
    \] 
    Let $E$ be the spectral family of $a$, $\eps > 0$ and $a_{\eps} := 
    \sum_{k = 1}^{m}\gl_{k}(E_{\gl_{k}} - E_{\gl_{k - 1}})$ such that $|a 
    - a_{\eps}| < \eps, \ E_{\gl_{m}} = I, \ E_{\gl_{0}} = 0$. There is a 
    unique $j$ such that $E_{\gl_{j}} - E_{\gl_{j - 1}} \in \gb$. Choose
    $p \in \gb$ such that $p(E_{\gl_{k}} - E_{\gl_{k - 1}}) = 0$ for $k
    \ne j$. Then
    \begin{eqnarray*}
	|pa| & ≤ & |pa_{\eps}| + |p(a - a_{\eps})|  \\
	& < & |\gl_{j}| + \eps  \\
	& ≤ & |\gt_{\gb}(a)| + 2\eps,
    \end{eqnarray*}
    so $|a|_{\gb} ≤ |\gt_{\gb}(a)|$. \\
    Thus, in general, $[\kcc]$ is an integer domain, but not a field.
\end{remark}
We can transfer the partial order of $\pr$ to $[\pr]$ by
\[
    \all \ P, Q \in \pr: \ ([P] ≤ [Q] \ :\llra \ \ex \ p \in \gb : \
    pP ≤ pQ).
\]
This is obviously well defined and satisfies the properties of partial
order. Let $P, Q \in \pr$ and let 
\[
    [P] \we_{\gb} [Q] := [P \we Q]. 
\]
This is well defined, and if $[E] ≤ [P], \ [E] ≤ [Q]$, then $[E] ≤ [P]
\we_{\gb} [Q]$: we have
$pE ≤ pP, \ pE ≤ pQ$ for some $p \in \gb$, hence $pE ≤ pP \we pQ$, i.e.
$[E] ≤ [P] \we_{\gb} [Q]$. Thus $[P] \we_{\gb} [Q]$ satisfies the
universal property of the minimum.\\
~\\
We return to the discussion of the rank over $\gb$. The following
result is decisive. It permits to generalize our proof that the
quasipoints of a finite \emph{factor} of type $\rm{I}$ are all atomic,
to a proof that the quasipoints of an arbitrary finite von Neumann algebra
of type $\rm{I}$ are all abelian.

\begin{proposition}\label{rk6}
    Let $\rr$ be a finite von Neumann algebra of type $\rm{I}_{n}$ and
   let $P, Q \in \pr$ be projections over $\gb \in \qc$ such that $rk_{\gb}(P)
   = rk_{\gb}(Q)$. Then $[P] ≤ [Q]$ implies $[P] = [Q]$.
\end{proposition}
\emph{Proof:} Replacing $P$ and $Q$ by $\scc(Q)P$ and $\scc(P)Q$,
respectively, we may assume that $P$ and $Q$ have equal central
support. Moreover, we can assume that $P ≤ Q$ holds. Then there are $p
\in \gb$ and orthogonal families $(E_{1},\ldots, E_{j}), \ (F_{1},\ldots, F_{j})$
of equivalent abelian projections such that 
\[
    E_{1} + \cdots + E_{j} = pP \ \textit{and} \ F_{1} + \cdots + F_{j} =
    pQ.
\]
Hence $pP = pQ$, since $\mnc$ is finite, and therefore $[P] =[Q]$. \ \
$\Box$\\
~\\
The rank over a quasipoint of the center shares also another property with 
the ordinary rank:

\begin{proposition}\label{rk6a}
    Let $\rr$ be a von Neumann algebra of type $\rm{I}_{n} \ (\nin)$ and let 
    $P \in \pbr$ be a  projection with $rk_{\gb}P < n$. Then $I - P \in \pbr$ and
    $rk_{\gb}(I - P) = n - rk_{\gb}P$.
\end{proposition}
\emph{Proof:} For an arbitrary projection $Q \in \rr$ let
\[
    c_{\kcc}(Q) := \bigvee \{p \in \pc \mid p ≤ Q \}.
\]
Then $c_{\kcc}(Q) = I - s_{\kcc}(I - Q)$ and $s_{\kcc}(I - P) = I - c_{\kcc}P$
implies that $I - P \in \pbr$ if and only if $c_{\kcc} \notin \gb$. But this
is equivalent to $rk_{\gb}P < n$. Let $k := rk_{\gb} $  and 
$m := rk_{\gb}(I - P)$. Since $\gb$ is a dual ideal, there is a $p \in \gb$ such
that 
\[
pP = E_{1} + \cdots + E_{k} \ \textit{and} \ p(I - P) = F_{1} + \cdots
+ F_{m}
\]
with abelian projections $E_{1}, \ldots E_{k}, \ldots, F_{1}, \ldots,
F_{m}$. This implies $pI = E_{1} + \ldots + E_{k} + F_{1} + \ldots +
F_{m}$, hence $k + m = n$. \ \ $\Box$\\

\begin{theorem}\label{rk7}
    Let $\rr$ be a von Neumann algebra of type $\rm{I}_{n} \ (\nin)$. 
    Then all quasipoints of $\rr$ are abelian.
\end{theorem}
\emph{Proof:} Let $\kcc$ be the center of $\rr$. Consider a quasipoint
$\frb$ of $\rr$ and the corresponding quasipoint $\gb := \frb \cap
\kcc$ of $\kcc$. Let 
\[
    r_{0} := \min \{ rk_{\gb}(P) \mid P \in \frb \}
\]
and choose $P_{0} \in \frb$ with $rk_{\gb}(P_{0}) = r_{0}$. Then we
obtain for an arbitrary $P \in \frb$:
\[
    r_{0} ≤ rk_{\gb}(P \we P_{0}) ≤ rk_{\gb}(P_{0}) = r_{0},
\]
i.e.
\[
    rk_{\gb}(P \we P_{0}) = rk_{\gb}(P_{0}).
\]
Hence, by proposition \ref{rk6}, $[P \we P_{0}] = [P_{0}]$. This
means
\[
    \all \ P \in \frb \ \ex \ p \in \gb : \ pP_{0} ≤ pP.
\] 
Since $rk_{\gb}(P_{0}) ≥ 1$, there is an abelian subprojection $E$ of
$P_{0}$ with $\scc(E) \in \gb$. If $P \in \frb$, then $pE ≤ pP_{0} ≤
pP$ for a suitable $p \in \gb$. So, in particular, $E \we P \ne 0$.
Hence, by the maximality of $\frb$, $E \in \frb$. \ \ $\Box$\\
~\\
Together with the results of section \ref{ac}, this theorem unveils
the structure of the Stone spectrum of a von Neumann algebra of type
$\rm{I}_{n} \ (\nin)$.

\section{Structure of the Stone spectrum}
\label{sr}

\begin{lemma}\label{sr1}
    If $\rr$ is of type $\rm{I}$, every abelian quasipoint $\frb$ of $\rr$
    contains an abelian projection with central support $I$.
\end{lemma}
\emph{Proof:} Let $E \in \frb$ be abelian and let $p := \scc(E)$.
Since $\rr$ is of type $\rm{I}$, there is an abelian $F \in \pr$ with 
central support $I$. Then
\[
    G := E + (I - p)F
\]
is abelian (\cite{kr2}), has central support $I$ and, as $E \in \frb$,
is contained in $\frb$. \ \ $\Box$\\
~\\
The proof shows that, for infinite-dimensional $\kcc$, an abelian quasipoint
of a von Neumann algebra of type $\rm{I}$ contains infinitely many abelian
projections with central support $I$.\\
~\\
We assume from now on that $\rr$ is a finite von Neumann algebra of
type $\rm{I}_{n}$. According to theorem \ref{rk7}, all quasipoints of 
$\rr$ are abelian.

\begin{lemma}\label{sr2}
    Let $E$ be an abelian projection over $\gb \in \qc$. Then
    there is exactly one $\frb \in \qr$ over $\gb$ that contains
    $E$.
\end{lemma}
\emph{Proof:} Let $\frb, \frb' \in \qr$ be quasipoints over $\gb$
that contain $E$. Since $s_{\kcc}(pE) = ps_{\kcc}(E)$ for 
all $p \in \pc$, the $E$-socles of $\frb$ and $\frb'$ are equal to
$\{ pE \mid p \in \gb \}$. Hence $\frb = \frb'$. \ \ $\Box$

\begin{corollary}\label{sr2a}
    Each fibre $\urb{\gz_{\kcc}}(\gb) \ (\gb \in \qc)$ is a discrete
    subspace of $\qc$ with respect to its relative topology.
\end{corollary}
Let $E$ be an abelian projection with central support $I$.
Then $E$ induces a section
\[
    \gs_{E} : \qc \to \qr.
\]
This follows directly from lemma \ref{sr2}: $\gs_{E}(\gb)$ is defined
to be the unique quasipoint over $\gb \in \qc$ that contains
$E$.\\
More generally, each abelian $E \in \pr$ induces, according to lemma
\ref{sr2}, a section 
\[
    \gs_{E} : \kQ_{\scc(E)}(\kcc) \to \qr.
\]
It is obvious that $\gz_{\kcc} \circ \gs_{E} = id_{\kQ_{\scc(E)}(\kcc)}$
holds.

\begin{lemma}\label{sr3}
    $\gs_{E}$ is continuous.
\end{lemma}
\emph{Proof:} Since $\qpr \cap \kQ_{E}(\rr) = \kQ_{P \we E}(\rr) =
\kQ_{pE}(\rr)$ for some $p \in \pc$, we have
\[
    \urb{\gs_{E}}(\qpr \cap \kQ_{E}(\rr)) = \kQ_{p}(\kcc) \cap
    \kQ_{\scc(E)}(\kcc),
\]
hence $\gs_{E}$ is continuous. \ \ $\Box$\\
~\\
The range of $\gs_{E}$ is $\kQ_{E}(\rr)$. It is the image of the compact set
$\kQ_{\scc(E)}(\kcc)$ by the continuous mapping $\gs_{E}$, hence
compact, too. This shows that $\qr$ is a \emph{locally compact space}.
Moreover, $\gz_{\kcc}$, restricted to $\kQ_{E}(\rr)$, is a
homeomorphism onto $\kQ_{\scc(E)}(\kcc)$, so $\gz_{\kcc} : \qr \to
\qc$ is a local homeomorphism. In other words, $\qr$ is a \emph{sheaf}
over $\qc$ with projection mapping $\gz_{\kcc}$ (\cite{war}, \cite{con}).\\
~\\
In a finite von Neumann algebra $\rr$, two projections $P, Q \in \rr$ 
are equivalent if and only if there is a \emph{unitary} $T \in \rr$
such that $Q = TPT^{\ast}$ (\cite{kr4}). The unitary group $\ur$
operates on $\qr$ by 
\[
    \all \ T \in \ur, \ \frb \in \qr : \ T.\frb := \{ TPT^{\ast} \mid 
    P \in \frb \}.
\]
Note that this operation is in accordance with the local operation of 
partial isometries. For, if $F = TET^{\ast}$ with $E \in \frb$ and
$T \in \ur$, $\gtt := TE$ is a partial isometry with $\gtt^{\ast}\gtt =
ET^{\ast}TE = E, \ \gtt\gtt^{\ast} = TET^{\ast} = F$ and, if $P ≤ E$, $\gtt P
\gtt^{\ast} = TEPET^{\ast} = TPT^{\ast}$. Therefore, $\gtt \frb_{E}
\gtt^{\ast} = (T.\frb)_{F}$, hence $\gtt_{\ast}\frb = T.\frb$.\\
~\\
It follows from proposition \ref{ac5} and theorem \ref{rk7} that $\ur$
operates \emph{transitively} on the fibres of $\gz_{\kcc}$. The
operation is, in general, not free. The \emph{isotropy group} of $\frb
\in \urb{\gz_{\kcc}}(\gb)$,
\[
    \ur_{\frb} := \{ T \in \ur \mid T.\frb = \frb \},
\]
can easily be determined.

\begin{proposition}\label{sr4}
    Let $\rr$ be a finite von Neumann algebra of type $\rm{I}_{n}$,
    $\frb \in \qr$ a quasipoint over $\gb \in \qc$ and $T \in \ur$. 
    Then the following properties of $T$ are equivalent:
    \begin{enumerate}
        \item  [(i)] $T.\frb = \frb$.
    
        \item  [(ii)] $TET^{\ast} \in \frb$ for some abelian $E \in
	\frb$.
    
        \item  [(iii)] $TET^{\ast} \in \frb$ for all abelian $E \in
	\frb$.
	
       \item  [(iv)] If $E \in \frb$ is abelian, there is some $p \in 
                  \gb$ such that $pTE = pET$, i.e. $[T]_{\gb}[E]_{\gb}
                   = [E]_{\gb}[T]_{\gb}$ for all abelian $E \in \frb$.
    \end{enumerate}
\end{proposition}
\emph{Proof:} $(iii) \lra (ii) \lra (i) \lra (iii)$ follows
immediately from lemma \ref{sr2}. If $E \in \frb$ is abelian such that
$TET^{\ast} \in \frb$, then $[E]_{\gb} = [TET^{\ast}]_{\gb}$ by
proposition \ref{rk6}. Thus $(ii)$ implies $(iv)$ and the converse is 
obvious. \ \ $\Box$\\
~\\
Because the action of $\ur$ is transitive on each fibre of
$\gz_{\kcc}$, the fibres are \emph{homogeneous spaces} $\ur /
\ur_{\frb}$, where $\frb$ is an arbitrary chosen element of
$\urb{\gz_{\kcc}(\gb)}$. Of course, this representation depends on the
choice of $\frb$, but $\ur_{\frb}, \ur_{\frb'} \ (\frb, \frb' \in
\urb{\gz_{\kcc}(\gb)})$ differ only by conjugation with a suitable
element of $\ur$. \\
~\\
We collect our results in the following 

\begin{theorem}\label{sr5}
    Let $\rr$ be a von Neumann algebra of type $\rm{I}_{n} \ (\nin)$
    with center $\kcc$. Then the Stone spectrum $\qr$ of $\rr$ is a
    locally compact space, the projection mapping $\gz_{\kcc} : \qr
    \to \qc$ is a local homeomorphism and, therefore, has discrete
    fibres. The unitary group $\ur$ of $\rr$ acts transitively on each
    fibre of $\gz_{\kcc}$. Therefore, each fibre
    $\urb{\gz_{\kcc}}(\gb)$ can be represented as a homogeneous space 
    $\ur / \ur_{\frb}$, where the isotropy group $\ur_{\frb}$ of $\frb
    \in \urb{\gz_{\kcc}}(\gb)$ is given by 
    \[
        \ur_{\frb} = \{ T \in \ur \mid [T]_{\gb}[E]_{\gb} =
	[E]_{\gb}[T]_{\gb} \ \textit{for all abelian} \ E \in \frb \}.
    \]
\end{theorem}

\section{Trace of a quasipoint on an abelian von Neumann subalgebra}
\label{tr}

The foregoing results can be used to answer the following question for
the special case of a von Neumann algebra $\rr$ of type $\rm{I}_{n} \ 
(\nin)$.\\
~\\
{\bf Problem:} Let $\frb$ be a quasipoint of a von Neumann
algebra $\rr$. Is there a maximal abelian von Neumann subalgebra $\mm$ 
of $\rr$ such that $\frb \cap \mm$ is a quasipoint of $\mm$?\\
$\frb \cap \mm$ is called the trace of $\frb$ on $\mm$.\\
~\\
This problem is of importance in the presheaf perspective of
observables (\cite{deg6}).\\
~\\
Let $\frb$ be a quasipoint of a von Neumann algebra of type $\rm{I}_{n}
\ (\nin)$ and let $\kcc$ be the center of $\rr$. Then $\gb := \frb
\cap \kcc$ is a quasipoint of $\kcc$. Choose any maximal abelian von
Neumann subalgebra $\mm'$ of $\rr$ and any quasipoint $\gga'$ of $\mm'$
that contains $\gb$. Moreover, let $\frb'$ be a quasipoint of $\rr$
that contains $\gga'$. $\frb$ and $\frb'$ are both quasipoints over
$\gb$, hence there is a $T \in \ur$ such that $\frb = T\frb' T^{\ast}$.
$\mm := T\mm' T^{\ast}$ is then a maximal abelian von Neumann
subalgebra of $\rr$ and $\gga := T\gga' T^{\ast}$ is a quasipoint of
$\mm$ that is contained in $\frb$. We have therefore proved:

\begin{proposition}\label{sr6}
    For each quasipoint $\frb$ of a finite von Neumann algebra of type
    $\rm{I}_{n}$ there is a maximal abelian von Neumann subalgebra
    $\mm$ of $\rr$ such that $\frb \cap \mm \in \qmm$.    
\end{proposition}
The next question that appears naturally is, whether $\frb \cap \mm$
is a quasipoint of $\mm$ for all maximal abelian von Neumann  subalgebras
of $\rr$.\\
If such a quasipoint would exist, it would induce, according to proposition
\ref{ac1}, a global section of the spectral presheaf of $\rr$, i.e. a 
family $(\gb_{\kaa})_{\kaa \in \frAr}$ of quasipoints $\gb_{\kaa} \in \qa$,
where $\frAr$ denotes the semilattice of all abelian von Neumann subalgebras
of $\rr$, with the property that 
\[
    \gb_{\kaa} = \gb_{\kbb} \cap \kaa \ \textit{for all} \ \kaa, \kbb 
    \in \frAr, \ \kaa \tm \kbb. 
\]
But, if $\rr$ has no direct summand of type $\rm{I}_{1}$ or
$\rm{I}_{2}$, an abstract form of the Kochen-Specker theorem
(\cite{doe}, \cite{ish3}) forbids global sections of the spectral presheaf.
We will prove directly that no quasipoint $\frb$ of a finite von Neumann 
algebra $\rr$ of type $\rm{I}_{n} \ (n \geq 2)$ has the property that 
$\frb \cap \mm \in \qmm$ for all maximal abelian von Neumann subalgebras of
$\rr$. In order to show this, it is convenient to represent $\rr$ as $\mnc$, the
$\kcc$-algebra of $(n, n)$-matrices with entries from the center $\kcc$
of $\rr$. $\mnc$ operates on $\ccn$ which we regard as a $\kcc$-module. 
(Since $\kcc$ is abelian, it does not matter whether we regard $\ccn$ as a
left or right $\kcc$-module.) The general structure behind this situation is
the theory of $C^{\ast}$-modules (\cite{la}).\\
If $a = (a_{1}, \ldots, a_{n})^{t} \in \ccn$, the $\kcc$-linear mapping 
\[
\begin{array}{cccc}
    E_{a} : & \ccn & \to & \ccn  \\
     & b & \tto & (b | a)a,
\end{array}
\]
where 
\[
    (b | a) := \sum_{k = 1}^{n}b_{k}a_{k}^{\ast}
\]
is the $\kcc$-valued ``scalar'' product on $\ccn$, is a projection if 
and only if $(a | a)$ is a projection in $\kcc$. A short calculation
shows that the projection $E_{a}$ is an abelian projection. 

\begin{remark}\label{sr7}
    All abelian projections of $\mnc$ are of the form $E_{a}$.
\end{remark}
\emph{Proof:} Choose a ``reference-projection'', say $E_{e_{1}}$,
where $e_{1} := (1, 0, \ldots, 0)^{t}$. Then $E_{e_{1}}$ has central
support $I$. If $E \in \mnc$ is an arbitrary abelian projection with
central support $p$, $E$ is equivalent to $pE_{e_{1}}$. Then there is 
a $T \in \kU(\mnc)$ such that 
\[
    E = TpE_{e_{1}}T^{\ast} = pE_{Te_{1}} = E_{pTe_{1}}.  \ \ \Box
\]
~\\ 
Now the property that $\frb \cap \mm \in \qmm$ for all maximal abelian 
von Neumann subalgebras of $\rr$, is obviously equivalent to the
property
\[
    \all \ P \in \pr : \ P \in \frb \ \textit{or} \ I - P \in \frb.
\]

\begin{lemma}\label{sr8}
    The property that $\frb \cap \mm \in \qm$ for all maximal abelian 
    von Neumann subalgebras of $\rr$, is equivalent to the property
    \[
    \all \ P_{1}, \ldots, P_{k} \in \pr : (P_{1} \vee \cdots \vee P_{k} 
    \in \frb \ \lra \ \ex \ i \leq k : \ P_{i} \in \frb).
    \]
\end{lemma}
\emph{Proof:} Let $P_{1} \vee  \cdots \vee P_{k} \in \frb$, but assume
that no $P_{i}$ belongs to $\frb$. Then, if the first property holds, 
$I - P_{1}, \ldots, I - P_{k} \in \frb$ and, therefore, 
\[
I - P_{1} \vee \cdots \vee P_{k} = (I - P_{1}) \we \cdots \we (I -
P_{k}) \in \frb,
\]
a contradiction. The converse is obvious. \ \ $\Box$

\begin{lemma}\label{sr9}
    $E_{a} = E_{a'}$ if and only if there is a unitary $u \in \kcc$ such that $a'
    = ua$. $u$ is unique iff $(a | a) = I$.
\end{lemma}
\emph{Proof:} If we consider the elements of $\kcc$ as continuous
functions on the Stone spectrum $\qc$ of $\kcc$, a unitary is a
continuous function $u : \qc \to S^{1}$. If $a, a' \in \{ b \in \ccn
\mid (b | b) \in \pc \}$ such
that $E_{a} = E_{a'}$, then $(a | a) = (a' | a')$, since $(a |
a)I_{n}$ is the central support of $E_{a}$. Moreover $a' = E_{a'}a' =
E_{a}a' = (a' | a)a$ and, symmetrically, $a = (a | a')a'$. Hence 
\[
    (a | a) = (a' | a') = (a' | a)(a' | a )^{\ast}(a | a),
\]
which implies $(a' | a)(a' | a )^{\ast} = 1$ on the support of $(a | a)$.
\[
    u(\gga) :=
    \begin{cases}
	(a' | a)(\gga)  & \ \textit{if} \ \gga \in S((a | a))  \\
	1                    & \ \textit{otherwise}
    \end{cases}
\]
defines a unitary $u : \qc \to S^{1}$ with $a' = ua$. The converse is 
obvious. \ \ $\Box$\\
~\\
Let $e := \frac{1}{\sqrt{n}}(1, \ldots, 1)^{t}$ and $E := E_{e}$, let 
further $e_{1}, \ldots, e_{n}$ be the ``unit vectors'' in $\ccn$ and
let $E_{k} := E_{e_{k}} \ (k = 1, \ldots, n)$. The projections $E,
E_{1}, \ldots, E_{n}$ are abelian  and have central support $I$. Since
the quasipoints $T \frb T^{\ast} \ (T \in \ur)$ have the property that
$T \frb T^{\ast} \cap \mm \in \qm$ for all maximal abelian von Neumann 
subalgebra $\mm$ if and only if $\frb$ has this property, we can assume that $E
\in \frb$. Because of $E_{1} + \cdots + E_{n} = I \in \frb$, we
conclude that $E_{k} \in \frb$ for some $k \leq n$. Since $E$ and
$E_{k}$ are abelian and have central support $I$, there is a $p \in
\frb \cap \kcc$ such that $pE_{k} = pE$. But, applying the 
foregoing lemma, this leads to a contradiction, since the components
$e_{kj}$ of $e_{k}$ are zero for $j \ne k$, while the corresponding
components of $e$ are equal to $\frac{1}{\sqrt{n}}$. Hence we have
proved 

\begin{proposition}\label{sr10}
    Let $\rr$ be a finite von Neumann algebra of type $\rm{I}_{n}$.
    The following properties are equivalent:
    \begin{enumerate}
        \item  [(i)] $n = 1$, i.e. $\rr$ is abelian,
    
        \item  [(ii)] $\ex \ \frb \in \qr \ \all \ P \in \pr: \ P \in 
	\frb \ \textit{or} \ I - P \in \frb$,   
    
        \item  [(iii)] $\all \ \frb \in \qr \ \all \ P \in \pr: \ P \in 
	\frb \ \textit{or} \ I - P \in \frb$.
    \end{enumerate}
\end{proposition}
Together with the abstract Kochen-Specker theorem we get

\begin{corollary}\label{sr11}
    If $\mm$ is a maximal abelian von Neumann subalgebra of a
    non-abelian von Neumann algebra $\rr$, there is a quasipoint $\frb$
    of $\rr$ such that $\frb \cap \mm \notin \qmm$.
\end{corollary}
If $\mm$ is a maximal abelian von Neumann subalgebra of a von Neumann 
algebra $\rr$, a quasipoint $\frb$ of $\rr$ is called \emph{admissible}
for $\mm$, if $\frb \cap \mm \in \qmm$. The foregoing results promise 
that the study of the sets 
\[
    \qr_{\mm} := \{\frb \in \qr \mid \frb \cap \mm \in \qmm \}
\]
and 
\[
    \frb_{\frAr} := \{\kaa \in \frAr \mid \frb \cap \mm \notin \qmm \}
\]    
will be an interesting task.

~\\
Hans F. de Groote\\
degroote@math.uni-frankfurt.de\\
Institut f\"{u}r Analysis und Mathematische Physik\\
FB Informatik und Mathematik\\
J.W. Goethe Universit\"{a}t\\
Frankfurt a.M.\\
Germany

\end{document}